\documentclass[twocolumn,aps,amsmath,nofootinbib]{revtex4}
\usepackage{graphicx}
\usepackage{amssymb}
\usepackage{color}
\definecolor{joerg}{rgb}{1.0,0.0,0.0}

\newcommand{\eq}{\begin{equation}}
\newcommand{\feq}{\end{equation}}
\newcommand{\hugenumber}{X}
\newcommand{\bl}{\left}
\newcommand{\br}{\right}

\begin{document}

\title{Modified dispersion relations from the renormalization group of gravity}

\author{F.~Girelli{}, S.~Liberati{}, R.~Percacci{}, C.~Rahmede{}}

\affiliation{{}SISSA and INFN, Sezione di Trieste \\
Via Beirut 2-4, 34014 Trieste, Italy}

\begin{abstract}
We show that the running of gravitational couplings, together with a suitable identification of the renormalization group scale can give rise to modified dispersion relations for massive particles. This result seems to be compatible with both the frameworks of effective field theory with Lorentz invariance violation and deformed special relativity.
The phenomenological consequences depend on which of the frameworks is assumed.
We discuss the nature and strength of the available constraints for both cases and show that in the case of Lorentz invariance violation, the theory would be strongly constrained.
\vspace*{1cm}\\
\end{abstract}
\maketitle

\section{Introduction}

The search for a quantum theory of gravity has been, for  more than half a century, a driving force in theoretical physics. Unfortunately, this research
has for a long time been frustrated by the intrinsic difficulty of testing theoretical predictions relevant at energy scales near  the Planck scale,
$M_{\rm P}=G^{-1/2}=1.22\cdot 10^{19}\; {\rm GeV}$ (we use units with $c=1$, $\hbar=1$). However this picture has recently been changed by the
realization that several (though not all) quantum gravity models seem to predict a departure from exact Lorentz invariance due to ultraviolet physics at
the Planck scale~\cite{LIV}. This could lead either to effective field theories (EFTs) characterized by Planck suppressed Lorentz violations for
elementary particles (see {\em e.g.}~\cite{Jacobson:2005bg}) or to some new physics where the Lorentz symmetry is deformed in order to include an
extra invariant scale (the Planck scale) apart from the speed of light~\cite{Amelino-Camelia:2000ge}. The latter framework is generally called
deformed special relativity (DSR).

Common to all these models is that they seem to predict a modified form of the  free particle dispersion relation, exhibiting extra momentum dependent terms, apart from the usual quadratic one occurring in the Lorentz invariant dispersion relation. In particular one most often
considers violations or deformations of the boost subgroup, leaving rotational invariance unaffected and leading to an expansion of the dispersion
relation in momentum dependent terms
\begin{eqnarray}
E^2&=&m^2+p^2  +F(p,\mu,M_{\rm P})
\nonumber\\
&=& m^2+p^2+ \sum_{n=1}^\infty \alpha_{n}(\mu,M_{\rm P})  \; p^n,
\label{eq:mod-disp}
\end{eqnarray}
where $p=\sqrt{||\vec p||^2}$
and $\mu$ is some particle physics mass scale.
In the following, we assume it to be equal to
$m$, the mass of the particle.

In the case of an EFT with Lorentz invariance violation (LIV), strong constraints on the coefficients $\alpha_n$ for  the cases
$n=1,2,3$ have been obtained~\cite{LIV}, and  there is some hope that LIV with $n=4$ will be constrained in the near future by forthcoming
experiments and improved observations~\cite{Jacobson:2005bg}.

In the case of deformed special relativity (DSR)~\cite{Amelino-Camelia:2000ge} constraints are more  uncertain, as we are still lacking a
satisfactory understanding of the theory in coordinate space and hence of the corresponding EFT. There are however conjectures about the
phenomenological implications of such a framework and some constraints have been tentatively provided (e. g., for the GZK threshold and
$n=3$~\cite{Heyman:2003hs}).

In this paper, we want to explore a mechanism that could lead to the emergence of such modified dispersion relations (MDRs), based on the idea that
the spacetime structure could be an emergent concept. If so, it would be natural to expect that at sufficiently high energies the effective spacetime
metric could become energy dependent. We shall see that sizeable effects could occur well below the Planck scale. It is interesting to note that such
a framework closely resembles that of emergent effective geometries and Lorentz symmetries characterizing many condensed matter systems (see {\em
e.g.}~\cite{LIV-Analog, EmSptm}). There, linearized perturbations propagate on Lorentzian geometries in the low energy (phononic) limit but do show
MDRs of the kind of Eq.~(\ref{eq:mod-disp}) as the perturbation's wavelength approaches the inter-molecular distance, or coherence length, below which
the background cannot be considered a continuum. This phenomenon can be seen as an energy dependent background metric interpolating between a purely
Lorentzian (and energy independent) form at very low energies, and a pre-geometric structure present at ultra-short length scales (sub-Planckian in
the quantum gravity scenario).

Albeit intriguing, the above described phenomenology of condensed matter systems is still an analogy after all. So one might wonder if  there is a
framework within known quantum gravity models that could naturally produce such an energy dependence of the metric. Some EFT models with this
property had been discussed in~\cite{Floreanini:1991cw}. There it was argued that in certain gauge theories of gravity with torsion, the
renormalization group (RG) flow of the couplings would produce a scale dependent metric. However, those calculations did not yield significant
effects below the Planck scale. More recently, other ideas in this direction have been advanced where an interpretation of the MDR used in
quantum gravity phenomenology was provided by arguing for an explicit dependence of the spacetime metric on the energy at which it is probed (see
{\em e.g.}~\cite{nick,Magueijo:2002xx,Eff-DSR} for alternative, but related, frameworks). For example in~\cite{Magueijo:2002xx}
Magueijo and Smolin proposed a generalization of DSR where the metric becomes energy dependent. In this paper we shall follow the general ideas of
~\cite{Reuter:1996cp}  where it was showed that within the context of an EFT of gravity, based on the conventional Einstein--Hilbert action with a
cosmological constant, it is indeed possible to derive an energy dependent metric from the RG flow of the couplings. From here, with some reasonable
assumptions, we will arrive at MDRs for the propagation of massive particles. We shall then analyze some constraints that can be cast on such quantum
gravity phenomenology.

\section{The RG of gravity}

There are many different viewpoints on what to expect from a quantum theory of gravity,
but a common feature to all is that they should reduce to general relativity 
at low energies.
In quantum field theory, a model that holds only in
a certain energy range is called an EFT.
The best--known example is the nonlinear sigma model description of mesons,
which is thought to be a low--energy approximation of QCD.
It is now well understood that Einstein's theory can be treated in the same way \cite{Donoghue:1994dn}.
Therefore, it is not necessary to regard Einstein's theory as a purely classical one:
graviton loop effects can be computed, as long as the corresponding momentum integrations
are cut off at some scale.

In any quantum field theory, whether fundamental or effective, the values of the coupling constants are not fixed but depend on the choice of a mass
scale $k$. Following Wilson's insight, an EFT describing physics at some energy scale $k$ is the result of functionally integrating all fluctuations
of the fields with momenta above $k$ (thus, $k$ can be regarded as an infrared cutoff). This results in an effective action $\Gamma_k$ for the
components of the fields with momenta lower than $k$. A priori in an EFT, $\Gamma_k$ will contain all the terms that are allowed by the symmetries of
the theory. Thus, in the case of gravity, it can be written as
\eq\label{effaction}
\Gamma_k(g_{\mu\nu})=\sum_i g_i(k) {\cal O}_i ,
\feq
where ${\cal O}_i$ are integrals of scalar functions of the metric and its derivatives,
and $g_i$ are coupling constants.
Clearly, this effective action depends on $k$;
the RG describes the way in which the couplings $g_i$
flow in dependence of $k$.
It is important to stress that at this general level the physical meaning of $k$ is not determined: for each physical process one will have to identify {the relevant RG scale}.
This is a crucial step requiring physical insight.

In this paper, we restrict ourselves to the sub-Planckian ($k^2\ll G^{-1}$) 
and small curvature ($k^2\gg R$) regime, where it is usually believed
that General Relativity is a good approximation
(for different views see {\it e.g.}~\cite{Sundrum:2003tb}).
We therefore assume the validity of the Einstein--Hilbert action, 
possibly with a cosmological constant:
\eq \Gamma_k\left[g\right]=\frac{1}{16\pi G_k}\int {\rm d}^4 x \sqrt g
\left(2\Lambda_k-R\right). \label{gAct} \feq

The running of the gravitational couplings can be computed inserting this action 
in an exact RG equation \cite{Wetterich:1992yh, Reuter:1996cp}. 
This approach can provide also nonperturbative information and has been applied 
in the search of a UV fixed point 
\cite{Lauscher:2001ya,Percacci:2005wu,Fischer:2006at}. 
Let us stress, however, that we do not need to commit ourselves to any specific
model of Planck scale physics, in particular we
do not assume the existence of an UV fixed point.

In the regime we are interested in, one obtains from the RG equation 
the following $\beta$-functions for Newton's constant and the cosmological constant
\cite{Reuter:1996cp}:
\begin{eqnarray}
\label{Grun1} k\frac{\partial}{\partial k}\left(\frac{1}{G_k}\right)&=& a\, k^2\\
\label{Lambdarun}k\frac{\partial\Lambda}{\partial k}&=& b\, G_k k^4
\end{eqnarray}
where $a$ and $b$ are some order one, positive coefficients
\footnote{ Note that if we allow for the presence of minimally coupled matter fields,
the form of equations (\ref{Grun1}) and (\ref{Lambdarun}) will not change but 
the values of $a$ and $b$ are affected \cite{Dou:1997fg}. We shall discuss the possible relevance of this fact in the conclusions.}.  
The $\beta$-functions for gravity have also been discussed 
in  different approaches (see {\em e.g.}~\cite{Shapiro:2000dz,Shapiro:2003ui,Shapiro:2004ch,Babic:2001vv,Guberina:2002wt,Babic:2004ev}).

The solutions to these ordinary differential equations are
\begin{eqnarray}
\frac{1}{G_k}&=& \frac{1}{G_{k_0}}+\frac{a}{2}\bl(k^2-k_0^2\br)\label{G_k}\label{Grun}\\
\Lambda_k &=& \Lambda_{k_0}+b\frac{ G_{k_0}}{4}\bl(k^4-k_0^4\br)\label{Lambda_k}\label{Lrun}.
\end{eqnarray}
Here $k_0$ is the scale at which the initial conditions are set. In fact, $G$ and $\Lambda$ are measured on quite different scales. The value
$G_{k_0}^{-1}=M_{\rm P}^2\approx 1.49 \cdot 10^{38} {\rm GeV}^2$ is measured to be the same from laboratory up to planetary distance scales, whereas for the
cosmological constant we have a value $\Lambda_{k_0}\approx 1.75 \cdot 10^{-123}M_{\rm P}^2$ measured at the Hubble scale $H_0\approx10^{-42}$ GeV~\cite{Perlmutter:1998np} .

From equatin (\ref{Grun}) we see that the running of $G$ is highly suppressed below $M_{\rm P}$ and hence will be neglected throughout. 
We see instead from Eq.~(\ref{Lrun}) that the running of $\Lambda$ becomes significant at energies of order $k_T\approx 10^{-31}M_{\rm P}\approx 10^{-3}$ eV or higher. 
(This corresponds to the ``turning point'', in the language of \cite{Reuter:2004nx}.) 
This significant running of the cosmological constant at relatively low scales 
will play a crucial role in our analysis.

We have not yet provided a prescription to determine $k_0$ as a function of some physical scale. Given however that there is no strong evidence for a present running of the cosmological constant at cosmological scales, we will assume, for the moment, that $k_0$ is placed below $k_T$ far enough in the infrared to be always negligible.
We shall check {\em a posteriori} that such an assumption is justified in the cases of our interest.

The equations of motion (EOM) at scale $k$ are obtained varying the effective  action with respect to the metric,
\eq\frac{\delta\Gamma_k}{\delta g_{\mu\nu}}=0.\feq
The solutions of the EOM at scale $k$ give the metric relevant for the physical process under consideration,
with the couplings evaluated at $k$.
In the theory with action (\ref{gAct}) the EOM are
\eq {R^{\mu}}_{\,\nu}\left[g_k\right]=\Lambda_k{\delta^{\mu}}_{\,\nu}.\label{EOM}\feq
Since ${R^{\mu}}_{\,\nu}\left[cg\right]=c^{-1}{R^{\mu}}_{\,\nu}\left(g\right)$ for any constant factor $c>0$,  equation (\ref{EOM}) can be rewritten as
\begin{eqnarray}
{ R^{\mu}}_{\,\nu}\left[g_{k_0}\right] &=&\Lambda_{k_0}{\delta^{\mu}}_{\,\nu} ={R^{\mu}}_{\,\nu}\left[\frac{\Lambda_{k}}{\Lambda_{k_0}}g_k\right],
\label{scal-deriv}
\end{eqnarray}
where we have used the coordinate independence of $\Lambda_k$. Therefore, for any solution of equation (\ref{EOM}) the inverse metric scales with the cosmological constant as~\cite{Lauscher:2005qz}
\begin{equation}
g^{\mu\nu}_k= \frac{\Lambda_{k}}{\Lambda_{k_0}}g^{\mu\nu}_{k_0}.
\label{eq:metr}
\end{equation}
We want now to analyze the consequences of such a scaling behaviour of the spacetime metric on the propagation of a free particle.

\section{Modified dispersion relations from a ``running" metric}

Starting from Eq.~(\ref{eq:metr}) we can derive a MDR by contracting it with the particle's four momentum and identifying $k$ with a function of the
momentum. In the presence of an effective cosmological constant the solution of the EOM cannot be flat space. However, we want to work in a regime
where the typical wavelength of the particle is much smaller than the characteristic curvature radius of spacetime, in our case $1/p\ll 1/\sqrt{R}\approx1/\sqrt{\Lambda_k}$. We can then
approximate $g^{\mu\nu}$ by a flat metric and equation (\ref{eq:metr}) just results in an overall scaling of the latter. Of course, in order to check that the above condition holds, we need to know the relation between $p$ and $k$. We shall check {\em a posteriori} that this is indeed the case for our choice of $k$.

A global rescaling of the metric can be eliminated by a choice of coordinates, but this can only be done at a particular scale. We choose
$g^{\mu\nu}_{k_0}=\eta^{\mu\nu}$ (the usual Minkowski metric $(1,-1,-1,-1)$) for any $k=k_0<k_T$.
 At scale $k>k_T$, we have the metric $g^{\mu\nu}_k$ defined by
\begin{equation}
g^{\mu\nu}_k= \frac{\Lambda_{k}}{\Lambda_{k_0}}\eta^{\mu\nu}; \label{eq:metr2}
\end{equation}
and contracting both sides of Eq.~(\ref{eq:metr2}) with the particle four momentum we then obtain
\begin{equation}\label{eq:metr3}
m^2
=\frac{\Lambda_{k}}{\Lambda_{k_0}}\eta^{\mu\nu} p_{\mu}p_{\nu} =\frac{\Lambda_{k}}{\Lambda_{k_0}}(E^2-p^2)\ ,
\end{equation}
where we have defined the mass to be $m^2=g_k^{\mu\nu}p_{\mu}p_{\nu}$ and $p_\mu=(E,-\vec p)$.
So, using Eq.~(\ref{Lrun}), one finally gets
\begin{eqnarray}\label{MDRgeneral}
E^2-p^2=\frac{\Lambda_{k_0}}{\Lambda_{k}}m^2=
\bl(1+\frac{b}{4}\hugenumber\frac{k^4}{M_{\rm P}^4}\br)^{-1}m^2\label{mdrgen}
\end{eqnarray}
where $\hugenumber={M_P^2}/{\Lambda_0}\approx 6\cdot 10^{122}$. 

In order to proceed further in our analysis we now need to clarify the relation between the RG scale $k$ and the particle momentum.
Assuming that rotational invariance is preserved, one can predict that for a free particle the RG scale $k$ will be generically determined by the modulus of the particle's three-momentum
$p:=\sqrt{\|\vec{p}\|^2}$ (or alternatively its energy given that they are practically the same, at first order, for high energy particles), its mass $m$, and possibly by the Planck scale.
As we expect that any deviation from standard physics should be Planck suppressed, we can then write the following ansatz
\eq
\label{kgeneral}
k=\frac{p^{\alpha}m^{\beta}}{M_{\rm P}^{\alpha+\beta-1}},
\feq
where $\alpha$ and $\beta$ are chosen to be positive integers. The above ansatz  is of course inspired by the standard framework adopted in most of the quantum gravity phenomenology literature (see {\em e.g.}~\cite{LIV}) and for any $\alpha\neq 0$ it will lead to dispersion relations characterized by higher order terms in the momentum of the particle suppressed by appropriate powers of the Planck mass.

For sufficiently low momenta the dispersion relations so obtained will take the form
\begin{equation}
E^2=p^2+m^2\left( 1-\frac{b}{4}X \left(\frac{m}{M_{\rm P}}\right)^{4\beta}
\left( \frac{p}{M_{\rm P}} \right)^{4\alpha}\right).
\label{eq:mostgenMDR}
\end{equation}
Note that due to the factor $m^2$ there is no modification of the dispersion relations for massless particles and that for a particle at rest $E=m$
as expected. Let us also emphasize that the above dispersion relation was derived assuming a point-like particle, as it is not clear at this stage
which quantities might enter in the relation between $k$ and the physical momentum for composite particles. For this reason, we shall in what follows focus on electrons/positrons.

In applications of the RG to high-energy physics, where one considers mainly scattering processes,  $k$ is usually identified with one of the
Mandelstam variables of the process, a Lorentz invariant combination of the incoming particle momenta. From this point of view the most obvious and
conservative choice would be to assume that $k$ is the unique Lorentz invariant function of the particle momentum, namely its mass. This would imply
$(\alpha,\beta)=(0,1)$. Of course from the perspective of this work this is an uninteresting choice, as it implies that, for a given particle type, $k$ is fixed
once for all: an ultra--high--energy particle would ``feel'' the same spacetime as one at rest. We shall then consider the case of  $\alpha\neq 0$.

Conversely one might wonder if there could be some strong motivation to rule out a priori the  mass dependence of relation (\ref{kgeneral}). One
possible argument can be based on the requirement that the natural condition $\Lambda_{k=M_{\rm P}}\approx M_{\rm P}^2$ holds. This implies that the
corresponding physical momenta will be $p=M_{\rm P} (M_{\rm P}/m)^{(\beta/\alpha)}$. 
The case $\beta=0$ is then the only one for which $k$ and $p$ would coincide at the Planck scale. Albeit appealing this feature of the $\beta=0$
class of models does not seem sufficient for excluding a priori the other kinds of dispersion relations. We shall hence, for the moment, consider all
the possible values of $\alpha$ and $\beta$ selecting them only on the basis of their phenomenological viability. However it is interesting to note
that, in the end, such analysis will indeed select for us a dispersion relation belonging to the $\beta=0$ class.

{Before dicussing the phenomenological viability of the above class of dispersion relations, 
it is perhaps important to stress that while the Planck scale dependence of  (\ref{eq:mostgenMDR}) 
does imply a departure from  standard GR at this scale, as generally expected, it does not necessarily conflict  with the possible existence of an UV fixed point~\cite{Lauscher:2001ya,Percacci:2005wu,Fischer:2006at}.  In fact the fixed--point action will almost certainly not be the 
Einstein--Hilbert action but could instead be some general diffeomorphism invariant 
action with extra degrees of freedom affecting local Lorentz invariance 
(like for example the well studied Einstein-{\ae}ther theory ~\cite{EA}). 
Furthermore, note that since the gravitons are massless, their propagation is not affected by (\ref{eq:mostgenMDR}). Thus, the presence of Planck suppressed terms in the propagators of massive particles and their detectability through carefully chosen experiments and observations \cite{LIV}
does not imply a sizable departure from standard GR at these sub-Planckian energies. 
It only means that the extra degrees of freedom characterizing the UV theory are weakly coupled to matter fields through Planck--suppressed interactions. }

\subsection{Phenomenological viability}

A good indicator of the phenomenological viability of the above class of dispersion relations is easily obtained  by considering when,  for
some choice of the parameters $\alpha$ and $\beta$, the Lorentz violating term becomes of the same order as the mass term, so that  the approximation
taken in order to derive Eq.~(\ref{eq:mostgenMDR}) breaks down.  This would indicate where the Planck--suppressed term starts introducing a running
mass term for the particle and hence producing a detectable phenomenology for example via threshold reactions. Results for such critical values of
the particle momentum for different choices of $\alpha$ and $\beta$ are given for the case of the electron  (assuming $b\approx O(1)$) in Table
\ref{tabular1}.
\begin{table}\nonumber
\begin{tabular}{|c|c|c|c|c|}\hline
 & $\alpha=1$ &$\alpha=2$ &$\alpha=3$ &$\alpha=4$ \\\hline
$\beta=0$ & $3\cdot 10^{-15}$ & $6.1$ & $7\cdot 10^{5}$ & $3\cdot 10^{8}$\\\hline $\beta=1$ & $7\cdot 10^{7}$ &  $9\cdot 10^{11}$&  $2
\cdot 10^{13}$& $1\cdot 10^{14}$ \\\hline $\beta=2$ & $2\cdot 10^{30}$ &  $1\cdot 10^{23}$&  $6\cdot 10^{20}$&  $4\cdot 10^{19}$\\\hline
\end{tabular}
\caption{The critical energies for order--one deviations from standard physics given in TeV for electrons/positrons at different combinations of
$\alpha$ and $\beta$.} \label{tabular1}
\end{table}
The dispersion relations (\ref{eq:mostgenMDR}) will be of course phenomenologically acceptable if the modifications to standard physics arise only for very high energy particles. 
However we do also have to ask that the critical value of the momentum is not too high so that the corresponding MDR might lead to observable effects and consequently be subject to observational constraints. 
For example, in the case of QED, the most interesting MDRs will be those for which observable effects are expected at TeV energies, as this is the scale of the most energetic QED
particles observed so far. Looking at Table \ref{tabular1} we see
that the only case of phenomenological interest seems to be $(\alpha,\beta)=(2,0)$. Higher values of $(\alpha,\beta)$ are not a priori incompatible
with observations, but at the moment lie beyond observational reach.

Before starting to consider the case   $(\alpha,\beta)=(2,0)$ let us note however that the cases with $\alpha=1$ are particularly interesting from
a theoretical point of view as they would lead to an MDR of the form
\begin{equation}
E^2=p^2+m^2+\eta_{\alpha=1}\,\frac {p^4}{M_{\rm P}^2}
\end{equation}
with $\eta_{\alpha=1}=-b/4\,X (m/M_{\rm P})^{2+4\beta}$.  
What is noticeable in our case is that the dimensionless coefficients $\eta$ do indeed contain, as conjectured (see {\em
e.g.}~\cite{Jacobson:2005bg}), powers of the small ratio $m/M_{\rm P}$. These are however not necessarily leading, in the present framework, to an
overwhelming suppression of the LIV term. In contrast, the presence of the huge numerical factor, $X$, that we inherited from the initial
conditions (the observed value of the cosmological constant on cosmological scales) basically allows us to rule out the most obvious case
$(\alpha,\beta)=(1,0)$ as this would lead to sizeable deviations from standard physics for any particle above $\approx 10^{-3}$ eV. If the observed $\Lambda_{k_0}$ contained also the contribution of some quintessence-like fields, the ``true"
cosmological constant at $k_0$ would be smaller hence leading to an even larger value of $X$.

The choice of parameters $(\alpha,\beta)=(2,0)$ gives \eq \label{scale2} k=\frac{p^2}{M_{\rm P}}, \feq and
\begin{equation}
E^2=m^2+p^2-\frac{b}{4}\hugenumber\,\frac{m^2 p^8}{M_{\rm P}^8}, \label{good-drel}
\end{equation}
which can be cast in the more suggestive form
\begin{equation}
E^2=m^2+p^2+\eta \frac{p^8}{M_{\rm P}^6}, \label{good-drel2}
\end{equation}
with $\eta=-({b}/{4})\hugenumber\,({m}/{M_{\rm P}})^2$.

Let us start noting that Eq.~(\ref{scale2}), together with Eq.~(\ref{Lambda_k}), implies that at sub-Planckian energies ($p\ll M_{\rm P}$) the de Broglie wavelength of the particle is always much smaller than the curvature radius of spacetime as we initially assumed. Furthermore it is also much smaller than the inverse of the RG parameter $k$. This can be interpreted as saying that the particle has a somewhat lower resolution in probing spacetime than naively expected.
We shall discuss this at length in the next section.  Passing to the dispersion relation
Eq.~(\ref{good-drel}) we already saw (see Table \ref{tabular1}) that it leads to  order--one deviations around 10 TeV. The reason for this lies again
in the presence of the huge numerical factor $X$ which is able to contrast the large Planck suppression. This feature makes the above dispersion
relation compatible with current (low energy) observations while at the same time amenable to experimental constraints via high-- energy astrophysics
observations of QED phenomena. From the experimental point of view, $\bl(\alpha,\beta\br)=\bl(2,0\br)$ is therefore the most interesting value. 
We shall argue now that theoretically it is the best motivated.

\subsection{Physical motivation for  the case $(\alpha,\beta)=(2,0)$}

In order to motivate physically the choice of the set of parameters $(\alpha,\beta)=(2,0)$ we shall start by addressing the question of the influence of the effective cutoff
on the fluctuations of the gravitational field, which is relevant for the
propagation of a free particle.
As mentioned above, $k$ marks the distinction between those modes
that are integrated over and those that have to be treated classically
in the resulting EFT.

As before, we assume that, in the absence of the particle, spacetime would be
effectively flat (or anyway would have a curvature much smaller than $m^2$).
The particle produces a disturbance in the gravitational field in the form
of local curvature and the fluctuations of the gravitational field
will be affected by this curvature.

We have to consider the specific dynamics describing the effect of the particle on the gravitational field, which we have assumed to be given by
Einstein's equations. Stripped of all indices, these equations tell us that the second derivatives of the metric, or the square of its first
derivatives, are of the order of $G\rho$, where $\rho$ is a typical component of $T_{\mu\nu}$. The solution depends on the distance from the
particle, and for a classical particle becomes singular near the origin. These issues do not arise when we take into account the quantum nature of
the particle. The position of the quantum particle cannot be determined with a precision greater than the de Broglie wavelength $\lambda=1/p$. For an
order of magnitude estimate, we can therefore spread the total energy and momentum $p$ in a box of size $1/p$, so the order of magnitude of the
diagonal components of the energy momentum tensor must be $\rho\approx p^4$. (We observe that this is also the vacuum energy density in a box of size
$1/p$.) Einstein's equations then give an estimate for the curvature
\eq
\bl(\partial g\br)^2\approx R\approx G\rho\approx \frac{p^4}{M_{\rm P}^2}.
\feq
Of course, the gravitational field will fall off away from the particle;
this equation gives just a characteristic value for the curvature very near the
position of the particle.
It says that
a particle of momentum $p$ excites Fourier modes of the gravitational
field with momenta up to $p^2/M_{\rm P}$.
Gravitational modes with higher momenta are essentially unaffected by the particle.
It is therefore natural to assume that in the EFT these are the modes which have
to be integrated over, meaning that the relevant cutoff is $k\approx p^2/M_{\rm P}\ll p$.
\footnote{Note that the same cutoff has been derived in other frameworks 
in \cite{Cohen:1998zx,Giddings:2005id}.}

It is interesting to observe that the same argument, applied to a charged particle in
an electromagnetic field, would yield a completely different result.
In fact, the order of magnitude of the charge and current density is $p^3$,
and from Maxwell's equations 
one gets an estimate
$
F\approx p^2.
$
Since $F$ has dimension of mass squared, we conclude that the characteristic momentum scale of the electromagnetic field generated by a particle of
momentum $p$ is $k\approx \sqrt{F}\approx p$.
This corresponds to the naive estimate $\alpha=1$.
Clearly, the different behavior is determined by the fact that
the coupling constant of electromagnetism is dimensionless,
while that of gravity has the dimension of area.

{In closing this section, we observe that in the case of a Friedmann--Lema\^\i tre--Robertson--Walker metric, our ansatz $k=\sqrt{G\rho}$ corresponds to the
Hubble scale~\footnote{Incidentally this confirms that our assumption of neglecting $k_0$ in Eq.~(\ref{Lrun}) was justified. In fact in this case $k_0\approx H_0\approx 10^{-33}$ eV which is definitely negligible for any particle with energies well above few meV.}. Interestingly this choice has also been advocated for applications to cosmology on the basis that the Bianchi-identity has to remain valid even in the presence of running coupling constants~\cite{Shapiro:2004ch,Guberina:2002wt,Babic:2004ev}. However, we stress that the framework presented here significantly differs from the one above. For example, in their argument the RG scale is assumed to be a function of the cosmological time $k=k(t)$. This would be incompatible with Eq.~(\ref{scal-deriv}) which explicitly relies on the coordinate independence of the cosmological constant.
{The same is valid for the models discussed in \cite{Reuter:2003ca,Reuter:2004nv}, where spherical symmetry of spacetime leads to a radial dependence of the couplings giving rise to a modified action from which also Brans--Dicke theory can be obtained.}
In our argument, the cutoff $k$ is determined by the properties of the object/apparatus that probes the spacetime metric, it does not depend {\em a priori} on the characteristic scale of the universe.}

\section{Physical interpretation and phenomenology}

The derived MDR has the form of a global scale transformation, so that a massless particle ends up probing the same spacetime no matter what energy
it has. However, there will be phenomenological consequences for massive particles. To make some precise experimental predictions, we need to choose
the framework in which we interpret this MDR. There are two main (inequivalent) candidates: EFT with LIV, or DSR. Both frameworks seem to be a priori
compatible with our MDR. The main difference between the two theories is in the way in which momenta add up and possibly in the spacetime
(non)commutativity.

\subsection{EFT with LIV}

The EFT framework has several advantages when discussing phenomenological consequences of LIV. Apart from being a well known and versatile framework
it is also able  to make sharp predictions as it allows to use the standard energy-momentum conservation and requires for its applicability just
locality and local spacetime translational invariance above some length scale. In this context, one can apply the usual QFT tools, bearing in mind
that the effective action will contain explicit Lorentz breaking terms, and hence cast constraints using experimental or observational tests. In
particular for Lorentz violations at order $O(p^3/M)$ and higher ({\em i.e.}~induced by nonrenormalizable operators of  mass dimension 5 or greater
in the action) the most appropriate tests are coming from high--energy astrophysical observations (see {\em e.g.}~\cite{Jacobson:2005bg}), as these
are among the highest energy phenomena we can access nowadays.

Not all of the above--cited astrophysical tests can be applied to our framework. In particular cumulative effects based on the propagation of photons
over cosmological distances~\cite{TOF,Jacobson:2005bg} are unavailable as no modification is induced in dispersion relations of massless particles.
Similarly some anomalous reactions, like the vacuum \v{C}erenkov $e^{\pm}\to e^{\pm}\gamma$ emission~\cite{Jacobson:2002hd}, are not allowed for
``subluminal" dispersion relations of the leptons (and unmodified photons) like the one we found in Eq.~(\ref{good-drel}). Finally one might consider
possible constraints coming from the shifting of normally allowed threshold reactions as the photon pair production, $\gamma\gamma\to e^+ e^-$,
or the GZK reaction, $p\gamma\to p \pi^0$ (with $\gamma$ a CMB photon, see {\em e.g.}~\cite{Aloisio:2000cm,Jacobson:2002hd}). Apart from the
previously mentioned caveats related to the application of our MDR to composite particles, such a route is again unfeasible within our framework. In
fact the analysis of such scattering reactions would require a new derivation of the relation between  the  RG parameter $k$ and the physical momenta
which (missing a better understanding of the theory) would require more arbitrary choices and assumptions on our side. Hence, given the above
theoretical and observational uncertainties, characterizing these scattering reactions  is beyond the scope of this paper.  We can however provide
very strong constraints on the MDR provided in Eq.~(\ref{good-drel}) by  considering the so called ``photon decay",
$\gamma\rightarrow e^+ e^-$, usually forbidden by momentum conservation, and the synchrotron emission.

\subsubsection{Photon decay}

Within an EFT framework the photon decay process becomes possible above a certain threshold energy once one has dispersion  relations violating
Lorentz invariance. This threshold energy is given by the minimal momentum of the incoming photon such that the decay could happen preserving energy--momentum conservation. Following the steps from \cite{Jacobson:2002hd} and referring to eqs. (\ref{MDRgeneral}) and (\ref{kgeneral}), one obtains for
the threshold momentum
\begin{equation}
p_{\rm th}= \bl(\frac{10^2}{b\hugenumber}\br)^{1/8}\bl(\frac{M_{\rm P}}{m_e}\br)^{\beta/2}M_{\rm P}\ .
\end{equation}
In the case $(\alpha,\beta)=(2,0)$, $b=1$ and using $m_e\approx 0.511$ MeV, the threshold energy for this process is $p_{\rm th}\approx 9.75$ TeV.
Moreover, following the steps in \cite{Jacobson:2005bg}, one can calculate the decay rate of the photon, which turns out to be extremely fast, so
much that a photon would not be able to propagate on any distance of astrophysical relevance. We then see that the observations of photons with
energies above $\approx 10$ TeV propagating on astrophysical distances allow us to put upper bounds on the coefficient $b$.

A strong constraint comes from the observation of high-energy $\gamma$-rays emitted by the Crab nebula \cite{Jacobson:2005bg}. The $\gamma$-ray
spectrum of this source is very well understood. It results from a high--energy wind of electrons (and possibly positrons) which leads to  a
combination of synchrotron emission and inverse Compton scattering of (mainly the synchrotron) photons. The inverse Compton $\gamma$-ray spectrum so
produced extends up to energies of at least 50 TeV. This implies that for these photons the threshold energy for their decay must be above 50 TeV. We
must have
\begin{equation}\label{gammadecay}
b\leq  \left(\frac{9.75\,\mbox{TeV}}{p_{\rm obs}}\right)^8,
\end{equation}
which for $p_{\rm obs}=50$ TeV gives a bound on $b$ of order 10$^{-6}$.
This is clearly a very strong constraint since $b$ is naturally of order one.

\subsubsection{Synchrotron radiation}

An even stronger constraint can be provided by the observation of high--energy synchrotron emission from the Crab nebula.
Cycling electrons in a magnetic field $B$ emit synchrotron radiation with a spectrum
that sharply cuts off at a frequency $\omega_c$ given by the formula
\begin{equation}
\omega_c=\frac{3}{2} eB\frac{\gamma^3(E)}{E}\, , \label{eq:opeaklv}
\end{equation}
where $\gamma(E)=(1-v^2(E))^{-1/2}$ and $v(E)$ is the electron's group velocity.
The formula (\ref{eq:opeaklv}) is based on the electron trajectory for
a given energy in a given magnetic field, the radiation produced by a given current,
and the relativistic relation between energy and velocity (see
\cite{Crab,Jacobson:2005bg} for a discussion about the validity
of this formula in EFT with LIV and a detailed derivation of the constraint).

The maximum synchrotron frequency $\omega_c^{\rm max}$ is obtained by maximizing
$\omega_c$ (\ref{eq:opeaklv}) with respect to the electron energy,
which amounts to maximizing $\gamma^3(E)/E$. Using the MDR (\ref{good-drel})
one can easily calculate the modified group velocity of the electron and
from that $\gamma(E)$. Then the maximization of the synchrotron frequency yields
\begin{equation}
\omega_c^{\rm max}=0.47 \, \frac{eB}{m_e}[-\eta\, (m_e/M_{\rm P})^6]^{-2/8}, \label{eq:opeaklv2}
\end{equation}
where $\eta=-b/4\,X\,(m_e/M_{\rm P})^2$. This maximum frequency is attained at the energy $E_{\rm
max}=(-m_e^2M_{\rm P}^6/35\eta)^{1/8}\approx 4.2\, b^{-1/8}$ TeV.

The rapid decay of synchrotron emission at frequencies larger than $\omega_c$ implies that most of the flux at a given frequency in a synchrotron
spectrum is due to electrons for which $\omega_c$ is above that frequency. Thus $\omega_c^{\rm max}$ must be greater than the maximum observed synchrotron emission
frequency $\omega_{\rm obs}$. This yields the constraint
\begin{equation}
b < \frac{4}{X}\left(\frac{M_{\rm P}}{m_e}\right)^8\left(\frac{0.47\, eB}{m_e\omega_{\rm obs}}\right)^{8/2}. \label{eq:synchcon}
\end{equation}
Using as in \cite{Crab} the observation of synchrotron emission from the Crab nebula up to energies of about 100 MeV, and a conservative estimate
of the magnetic field of 0.6 mG (this is the largest proposed value which yields the weakest constraint) we then infer\footnote{Note that possible
complications related to different MDRs between electrons and positrons in EFT with LIV are not present here as the breakdown of Lorentz invariance is
a geometric effect in our framework and as such will not distinguish between leptons of same mass.}  that $b\lesssim 2\times10^{-22}$. This constraint is so
strong that one has to conclude that dispersion relations like (\ref{good-drel}) for leptons are ruled out by current astrophysical observations
within an EFT framework~\footnote{One might wonder, given the strength of the synchrotron constraint for $(\alpha,\beta)=(2,0)$, if this might also help constraining cases  which require higher, but not totally unreasonable, critical energies like the cases $(\alpha,\beta)=(1,1)$ and $(\alpha,\beta)=(3,0)$. Unfortunately it is easy to check that  this is not the case. For example, for $(\alpha,\beta)=(1,1)$, the synchrotron constraint is just $b\lesssim 10^{17}$ and for $(\alpha,\beta)=(3,0)$ it is $b\lesssim 10^{29}$.
}.

\subsection{Deformed Special Relativity}

It is less easy to provide experimental constraints in the  DSR  framework since this theory still lacks a clear understanding. In particular in this
approach spacetime is in general noncommutative, although this non-commutativity can be absorbed by a coordinate transformation in phase space
\cite{GKKL}. Hence the form of DSR in coordinate space is still debated. In momentum space DSR is described by a deformation of the Lorentz symmetry
so that the  $p_\mu$  carries a nonlinear representation of the Lorentz group.  This nonlinear representation can be constructed from a linear
representation carried by a ``platonic'' momentum $\pi_\mu$ via a non-linear map $U$ so that $\pi_\mu=U^{-1}(p_\mu)$. The $\pi_\mu$ add linearly which implies a nonlinear addition for $p_\mu$.
It is still unclear why the latter nonlinear momenta should be the physically measured ones and there is a rich literature now devoted to the
interpretation of DSR in momentum space. Such an interpretation should hopefully clarify some of the problems pointed out in the recent
literature~\cite{Schutzhold:2003yp}.

 DSR can also be seen as a natural framework for our results if one recalls its interpretation as a new  measurement theory \cite{Eff-DSR}.
Generally we measure the momentum $\pi^\mu$ of a particle in a given reference frame described by a tetrad field ${e^\alpha}_\mu$, so that the 
measurement outcomes are $p^\alpha={e^\alpha}_\mu\pi^\mu$.  If the metric is endowed with quantum gravity fluctuations, 
the theory of measurement will imply an averaging procedure at some given energy scale possibly provided by the test particle. This can naturally
lead to an energy dependent tetrad field and hence to a nonlinear relation between the measurement outcomes $p^\alpha$ and the particle momentum
$\pi^\mu$~\cite{Eff-DSR}. This is precisely what happens in Eq.~(\ref{good-drel}).  At the scale $k_0$, since the metric is just the Minkowski
metric, $\pi^\mu$ and $p^\alpha$ coincide and transform linearly under Lorentz transformations. As soon as we get to higher energies the metric, and
hence the tetrad field, becomes momentum dependent: $g^{\mu\nu}(p)=\eta ^{\alpha\beta}{e ^{\mu}}_{\alpha}(p){e^{\nu}}_{\beta}(p)$. The nonlinear
relation between $\pi^\mu$ and $p^\mu$ is then given by \eq \pi_\mu=\sqrt{\frac{\Lambda_{k}}{\Lambda_{k_0}}}p_\mu= \bl(\sqrt{1+\frac{b}{4}\hugenumber
\frac{p^8}{M_{\rm P}^8}}\br)p_\mu . \feq Clearly, if $\pi^\mu$ undergoes linear Lorentz transformations, $p^\mu$ is transformed nonlinearly,
which is characteristic of the usual DSR framework.

With regard to phenomenological constraints, the main experimental prediction of DSR concerns  $\gamma$-ray  bursts \cite{Amelino-Camelia:1999pm},
but photons are not affected in our framework. Even worse, in this case we cannot resort to anomalous (normally forbidden) threshold reactions as the
latter are not allowed in DSR either. The reason for this is simply that a kinematically forbidden reaction in the ``platonic" variables $\pi_\mu$
cannot be made viable just via a nonlinear redefinition of momenta. There have been attempts to consider constraints provided by shifts of normally
allowed threshold reactions~\cite{Heyman:2003hs}. We note here that for such reactions the possible constraints are strongly dependent not just on
kinematical considerations, but also on reaction rates which require some working framework for their derivation. Hence, missing a field theory
description of DSR we cannot safely pose such constraints. Similar considerations hold for constraints based on the synchrotron emission~\cite{Crab}.

\section{Conclusion}

We have shown how the RG of gravity could lead to MDRs for massive minimally coupled particles due to the effects of quantum fluctuations. We have
argued that for a free particle the most plausible identification of the cutoff is $k=p^2/M_P$, where $p$ is the particle three momentum, leading to sizeable 
effects in the region of $p\approx$ 10 TeV for QED processes. To do this, we had to make several assumptions: we assumed the validity of Einstein's theory of
gravity from cosmological to particle physics scales ({\it i.e.}~still much below the Planck energy), we assumed that the cutoff is a function of three momentum squared rather than four momentum squared, and we took for granted the value $\Lambda_0\approx 10^{-85}\; {\rm GeV}^2$ for the cosmological constant at cosmological scales. At no point was it necessary to assume the existence of a gravitational fixed point, as this concerns physics at or beyond the Planck scale.

Another implicit assumption was the identification of the components
of the four momentum $p_\mu=(E,-\vec p)$, rather than $p^\mu=(E,\vec p)$. The two identifications are not compatible if the metric is scale
dependent. Had we chosen $p^\mu=g_k^{\mu\nu}p_\nu=(E,\vec p)$, we would have obtained a MDR of the type
\begin{equation}
E^2=p^2+ m^2 \bl(1+\frac{b}{4}\frac{\hugenumber k^4}{M_{\rm P}^4}\br).
\end{equation}
The main difference with respect to Eq.~(\ref{good-drel}) is the positive sign in front of the correction. From an LIV EFT perspective, the immediate consequence is
that the previously discussed synchrotron bound does not apply, so this MDR is not as constrained as the previous one (Eq.~(\ref{good-drel})). There is no photon decay either, however a vacuum \v{C}erenkov effect may occur~\cite{LIV}. 
It can be shown that the latter can cast constraints on $b$ of the same order as the photon decay case (cf Eq.
(\ref{gammadecay})).
Thus this case would also be quite constrained in the LIV interpretation. Note that a similar change in the sign of the LIV term in the MDR could
also be due to a change in the sign of the coefficient $b$ which can be induced by considering in the RG analysis also minimally coupled fermion
fields~\cite{Dou:1997fg}.
Different initial assumptions may lead to effects that are either too strong to be compatible with current data, or too small to be detectable in the
foreseeable future, or, hopefully, they might produce some interesting phenomenology.

The RG has been applied in an LIV context in \cite{Nielsen} whose authors proved that Lorentz
invariance can arise as a low--energy symmetry in an otherwise non--Lorentz invariant
theory. Since our MDRs reduce to the standard ones at sufficiently low energies,
our results are in agreement with theirs on this point, even though the formalism
is quite different.

From the theoretical point of view, we cannot say at this point if the RG approach to gravity prefers EFT with LIV or DSR. Let us stress that our model is not a priori equivalent to any of the above frameworks.  In fact in the case of EFT with LIV one assumes the
existence of some aether field which allows one to construct LIV operators in the matter Lagrangians.  However in our case the departure from standard
special relativistic dynamics of matter is induced uniquely via the $k$ dependence of the background metric and the fact that $k$ is chosen not to be
a Lorentz invariant. This can be seen as a special case of EFT with an aether field but it is not equivalent to it. (Note that if the aether
field is taken to be dynamical then it will affect the RG flow, but this will be equivalent to the presence of an extra matter field.) Similarly
we cannot a priori completely identify our framework with a DSR one since we cannot say at this stage if our effective geometry will  also be
accompanied by some alternative rule for the addition of momenta.

Which of the above possibilities would be actually realized within our framework could  be probably assessed only after gaining a better
understanding of EFT on running geometries, something that is still lacking at this time. Following the intuition that comes from the analogue
models (where the underlying microscopic physics indeed violates  Lorentz invariance), one would probably  need to have some deeper understanding of the physics above the Planck scale (quantum gravity regime)
to be able to distinguish between the two. 
In this sense the phenomenological analysis we have performed has to be taken as a first try aimed at seeing what constraints could be cast once this
discrimination is done.

\section*{Acknowledgements}
The authors wish to thank David Mattingly and Lorenzo Sindoni for constructive criticisms.


\end{document}